\documentclass[a4paper,11pt]{article}
\usepackage{aaskaiid}
\usepackage[subpreambles=true]{standalone}
\usepackage{import}
\usepackage{xcolor}

\newcommand\arcsec{\hbox{$^{\prime\prime}$}}
\usepackage{orcidlink}
\title{Compact radio galaxies: the case of FR0s}
\ShortTitle{FR0 radio galaxies}

\author[1]{R. D. Baldi\orcidlink{0000-0002-1824-0411}}
\ShortName{Baldi et al.} 
\author[2]{A. Capetti\orcidlink{0000-0003-3684-4275}}
\author[1,3]{G. Giovannini\orcidlink{0000-0003-4916-6362}}
\author[4]{S. Amarantidis\orcidlink{0000-0001-7948-5714}} 
\author[1]{M. Brienza \orcidlink{0000-0003-4120-9970}}
\author[5]{G. Bruni \orcidlink{0000-0002-5182-6289}}
\author[6]{J. Chilufya\orcidlink{0000-0002-7420-1505}}
\author[2]{A. Costa\orcidlink{0009-0006-4561-4446}}
\author[3,1]{M. Gitti\orcidlink{0000-0002-0843-3009}}
\author[7]{D. V. Lal\orcidlink{0000-0001-5470-305X}}
\author[1]{G. Migliori\orcidlink{0000-0003-0216-8053}}
\author[8]{J. Moldon\orcidlink{0000-0002-8079-7608}}
\author[1]{M. Orienti\orcidlink{0000-0003-4470-7094}}
\author[5]{F. Panessa\orcidlink{0000-0003-0543-3617}}
\author[1]{I. Prandoni\orcidlink{0000-0001-9680-7092}}
\author[8]{M. Puig-Subirà\orcidlink{0000-0001-8955-7574}}
\author[9]{R. M. Samir} 
\author[10]{S. Shabala\orcidlink{0000-0001-5064-0493}}
\author[11]{F. Shankar\orcidlink{0000-0001-8973-5051}}
\author[1]{C. Spingola\orcidlink{0000-0002-2231-6861}}
\author[12]{F. Tavecchio\orcidlink{0000-0003-0256-0995}}
\author[3,1]{F. Ubertosi\orcidlink{0000-0001-5338-4472}}
\author[13]{B. Vaidya\orcidlink{0000-0001-5424-0059}}

\small
\affiliation[1]{INAF - Istituto di Radioastronomia, via Gobetti 101, 40129 Bologna, Italy}
\emailAdd{ranieri.baldi@inaf.it}
\affiliation[2]{INAF - Osservatorio Astrofisico di Torino, Strada Osservatorio 20, 10025 Pino Torinese, Italy}
\affiliation[3]{Dipartimento di Fisica e Astronomia, Universita' di Bologna, Via P. Gobetti 93, I-40129 Bologna, Italy}
\affiliation[4]{Institut de Radioastronomie Millimétrique (IRAM), Avenida Divina Pastora 7, Local 20, E-18012, Granada, Spain}
\affiliation[5]{INAF - Istituto di Astrofisica e Planetologia Spaziali, via del Fosso del Cavaliere 100, Roma, I-00133, Italy}
\affiliation[6]{Centre for Astrophysics Research, Department of Physics, Astronomy and Mathematics, University of Hertfordshire, Hatfield AL10 9AB, UK}
\affiliation[7]{National Centre for Radio Astrophysics - Tata Institute of Fundamental Research Post Box 3, Ganeshkhind P.O., Pune 411007, India}
\affiliation[8]{Instituto de Astrofísica de Andalucía, Consejo Superior de Investigaciones Científicas (CSIC), Glorieta de la Astronomía s/n, E-18008 Granada, Spain}
\affiliation[9]{Astronomy Department, National Research Institute of Astronomy and Geophysics (NRIAG), EL Marsad Street 1, Helwan, Cairo, Egypt}
\affiliation[10]{School of Natural Sciences, University of Tasmania, Private Bag 37, Hobart, TAS, 7001, Australia}
\affiliation[11]{School of Physics and Astronomy, University of Southampton, Highfield, Southampton, SO17 1BJ, UK}
\affiliation[12]{INAF – Osservatorio Astronomico di Brera, Via E. Bianchi 46, I-23807 Merate, Italy}
\affiliation[13]{Disciple of Astronomy, Astrophysics and Space Engineering, 
Indian Institute of Technology Indore Khandwa Road, Simrol, Indore 453552}

\abstract{Fanaroff–Riley type 0 (FR0) radio galaxies, a newly-identified and abundant population of low-power radio-loud active galactic nuclei, present a significant challenge to our understanding of radio galaxy evolution. Unlike their more extended FRI and FRII counterparts, FR0s are characterized by compact (from pc to a few kpc) radio morphologies with weak or absent large-scale jets, despite having optical host galaxy properties and central black hole masses similar to classical, more powerful radio galaxies. Their compactness and prevalence suggest they may represent an early stage of radio galaxy evolution or indicate a fundamentally different accretion and ejection mechanism. The Square Kilometre Array (SKA), with its continuum survey and VLBI capabilities, offers a transformative opportunity to study FR0 radio galaxies at unprecedented angular resolution and sensitivity. Milliarcsecond (mas) angular resolution VLBI observations will probe their parsec-scale structure, providing critical insights into the nature of their jets, accretion-ejection physics, and the interplay between nuclear activity and the surrounding environment. Continuum and polarization surveys will enable a systematic study of their population properties, distribution, and radio spectra across a wide range of redshifts in relation to other populations of radio galaxies.}

\begin{document}
\newcommand{\actaa}{Acta Astron.} 
\newcommand{\araa}{Annu. Rev. Astron. Astrophys.} 
\newcommand{\aar}{Astron. Astrophys. Rev.} 
\newcommand{\ab}{Astrobiol.} 
\newcommand{\aj}{Astron. J.} 
\newcommand{\apj}{Astrophys. J.} 
\newcommand{\apjl}{Astrophys. J. Lett.} 
\newcommand{\apjs}{Astrophys. J. Suppl. Ser.} 
\newcommand{\ao}{Appl. Opt.} 
\newcommand{\apss}{Astrophys. Space Sci.} 
\newcommand{\aap}{Astron. Astrophys.} 
\newcommand{\aapr}{Astron. Astrophys. Rev.} 
\newcommand{\aaps}{Astron. Astrophys. Suppl.} 
\newcommand{\baas}{Bull. Am. Astron. Soc.} 
\newcommand{\caa}{Chinese Astron. Astrophys.} 
\newcommand{\cjaa}{Chinese J. Astron. Astrophys.} 
\newcommand{\cqg}{Class. Quantum Gravity} 
\newcommand{\gal}{Galaxies} 
\newcommand{\gca}{Geochim. Cosmochim. Acta} 
\newcommand{\icarus}{Icarus} 
\newcommand{\jcap}{J. Cosmol. Astropart. Phys.} 
\newcommand{\jgr}{J. Geophys. Res.} 
\newcommand{\jgrp}{J. Geophys. Res.: Planets} 
\newcommand{\jqsrt}{J. Quant. Spectrosc. Radiat. Transf.} 
\newcommand{\memsai}{Mem. Soc. Astron. Italiana} 
\newcommand{\mnras}{Mon. Not. R. Astron. Soc.} 
\newcommand{\nat}{Nature} 
\newcommand{\nastro}{Nat. Astron.} 
\newcommand{\ncomms}{Nat. Commun.} 
\newcommand{\nphys}{Nat. Phys.} 
\newcommand{\na}{New Astron.} 
\newcommand{\nar}{New Astron. Rev.} 
\newcommand{\physrep}{Phys. Rep.} 
\newcommand{\pra}{Phys. Rev. A} 
\newcommand{\prb}{Phys. Rev. B} 
\newcommand{\prc}{Phys. Rev. C} 
\newcommand{\prd}{Phys. Rev. D} 
\newcommand{\pre}{Phys. Rev. E} 
\newcommand{\prl}{Phys. Rev. Lett.} 
\newcommand{\psj}{Planet. Sci. J.} 
\newcommand{\planss}{Planet. Space Sci.} 
\newcommand{\pnas}{Proc. Natl Acad. Sci. USA} 
\newcommand{\procspie}{Proc. SPIE} 
\newcommand{\pasa}{Publ. Astron. Soc. Aust.} 
\newcommand{\pasj}{Publ. Astron. Soc. Jpn} 
\newcommand{\pasp}{Publ. Astron. Soc. Pac.} 
\newcommand{\rmxaa}{Rev. Mexicana Astron. Astrofis.} 
\newcommand{\sci}{Science} 
\newcommand{\sciadv}{Sci. Adv.} 
\newcommand{\solphys}{Sol. Phys.} 
\newcommand{\sovast}{Soviet Ast.} 
\newcommand{\ssr}{Space Sci. Rev.} 
\newcommand{\uni}{Universe} 

\maketitle

\section{Introduction}
\begin{figure}
   \centering
   \includegraphics[width=0.8\hsize]{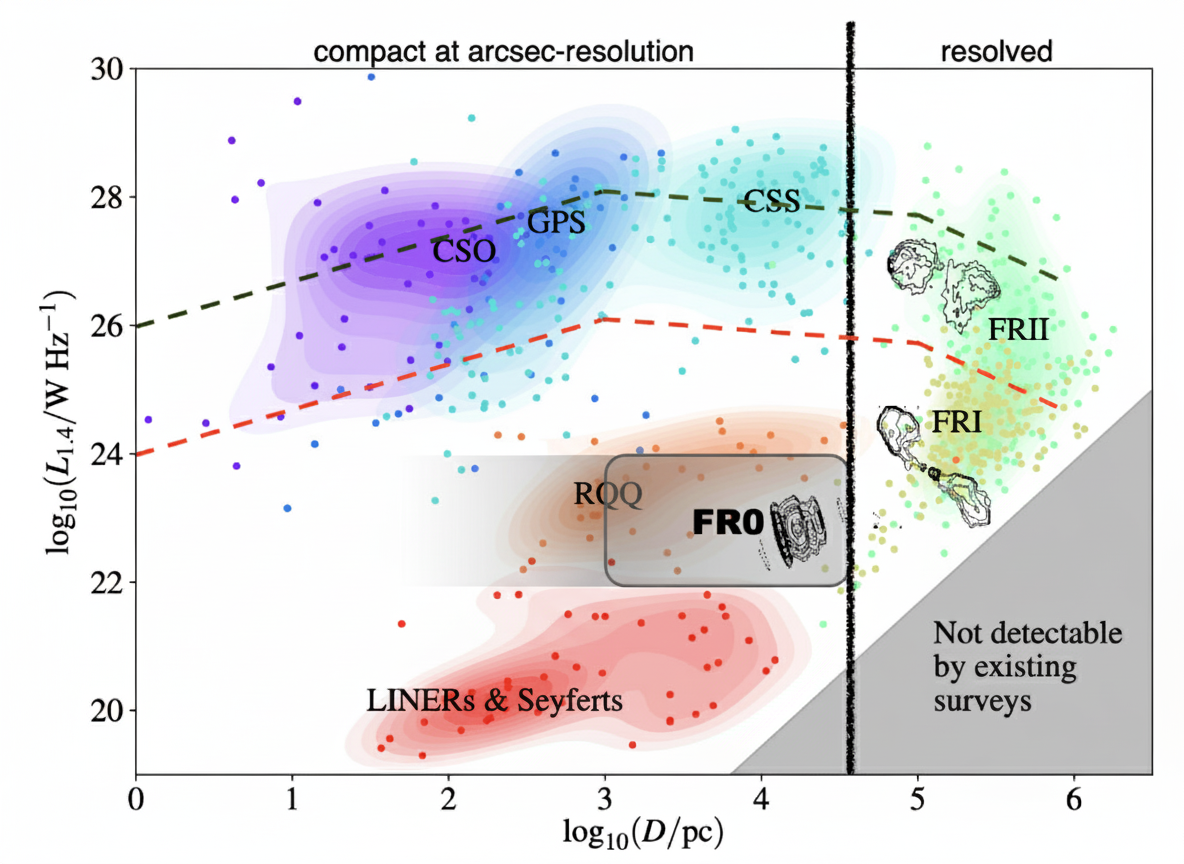}
        \caption{Radio power/linear-size plot (P-D diagram) for different types of RL and radio-quiet (RQ) AGN, adapted from plots presented by \citet{hardcastle20,baldi23}. Points show individual objects and coloured contours represent a smoothed estimator of source density. FR0s occupy the compact, low-power AGN region ($\sim$-1-10 kpc), represented by the box fading at smaller sizes. The diagram also includes young RGs (CSO, GPS and CSS); adult RGs (FRI, FRII); and RQ/intermediate AGN (LINERs, Seyferts, RQ Quasars). Classical evolutionary tracks (CSO, GPS, CSS $\rightarrow$ FRI/II) are also shown with red and dark-green dashed lines \citep{an12}. The vertical line roughly represents the separation between resolved and unresolved/compact sources based on arcsec angular resolution.  Three different morphologies are also displayed: 3C285 as example of FRII, Centaurus A (inner lobes) as example of FRI and pc-scale jet structure of a FR0 \citep{baldi24}.}
        \label{diagram}
\end{figure}

Radio galaxies (RGs) consist in a class of radio-loud (RL) active galactic nuclei (AGN) characterized by relativistic jets launched from accreting supermassive black holes (BHs). Historically, extended jetted sources were classified as Fanaroff–Riley type~I and II (FRI and FRII, \citealt{fanaroff74}), based on their radio morphologies (spanning up to Mpc scales) and jet powers. FRIs exhibit less powerful, core-brightened structures, whereas FRIIs are edge-brightened and lobe-dominated (Fig.~\ref{diagram}). This classification also correlates with differences in the host galaxy properties, clustering, and accretion states (e.g., \citealt{zirbel95,heckman14,mingo22}). However, recent multi-wavelength surveys are revealing a more complex picture than this traditional dichotomy suggests (e.g., \citealt{urry95,padovani17b}).

Among the most significant discoveries within this ever-evolving RLAGN framework is the identification of a class of local, compact, low-power RGs termed Fanaroff–Riley type 0 (FR0, \citealt{ghisellini11,baldi16}, Fig.~\ref{diagram}). These sources lack extended kpc-scale radio emission and constitute the most numerous population of low-luminosity ($L < 10^{24}$ W Hz$^{-1}$) RGs in the local Universe ($z < 0.05$), outnumbering FRIs and FRIIs by a factor of $\gtrsim 5$ \citep{baldi23}. Their radio emission is typically unresolved with arcsecond resolution, core-dominated, and generally features flat spectra (F$_{\rm \nu} \sim \nu^{-\alpha}$ with $\alpha <$ 0.5) in the GHz domain. High-resolution observations often detect pc-scale jets that quickly fade or become disrupted on sub-kiloparsec scales \citep{baldi19,cheng18,cheng21,giovannini23}. 

The origin of FR0s remains an open question in AGN physics \citep{baldi23}. Their sheer abundance compared to FRIs presents a fundamental evolutionary problem and suggests a phase of inefficient jet expansion. Proposed theoretical scenarios include a static model, where the jets are intrinsically weak, and a dynamic model, where FR0s represent short-lived or intermittent AGN phases \citep{sadler14}. The evolution of low-power compact sources like FR0s is also poorly understood, especially when compared to their more powerful counterparts, Compact Symmetric Objects (CSO), GHz-Peaked Spectrum sources (GPS), and Compact Steep-Spectrum sources (CSS), which are generally interpreted as FRI/II progenitors (e.g., \citealt{an12}). Specifically, standard evolutionary tracks in the radio power-linear size ($P-D$, Fig.~\ref{diagram}) diagram fail to adequately explain this numerous population of local, low-luminosity, compact RLAGN \citep{hardcastle20}. 

Multi-band studies have established that FR0s reside in red (quiescent) massive elliptical galaxies with BH masses ranging from $10^{7.5}$–$10^9$ $M_{\odot}$ \citep{sadler14} and \citep{baldi18}. Their optical and X-ray nuclei are consistent with radiatively inefficient accretion flows, which are typical of low-excitation RGs (LERGs) \citep{torresi18}. Furthermore, their compact jets can efficiently couple with the interstellar medium (ISM), contributing to radio-mode feedback \citep{mukherjee16,hardcastle19,igo25}. They can also emit at very high-energy bands (see SKA chapter on high-energy RGs from \citealt{Bruni01.2026.SKA}), leading to the idea that they may contribute to the extragalactic $\gamma$-ray background and be cosmic neutrino emitters 
\citep{grandi16,tavecchio18,baldi19rev,stecker19,palyia21}.

Whether FR0s represent the youth of a RG or signify a specific mode of the RLAGN phenomenon remains unclear, especially since current radio surveys are biased toward luminous or extended sources. The Square Kilometre Array (SKA) will be transformative: the unprecedented combination of high sensitivity, high resolution, and high-image fidelity will enable a comprehensive census of the FR0 population across cosmic time. This chapter (together with the SKA chapter on RGs from \citealt{Hardcastle01.2026.SKA}) outlines our current understanding of FR0s and discusses how SKA will address the open questions concerning their origin, duty cycle, and impact on galaxy evolution.

\section{Current radio-band  view of FR0s}

The radio properties of FR0 RGs, primarily limited to the local Universe (z $<$ 0.1), have been extensively investigated over the last decade using a wide range of radio observational facilities.

At low frequencies ($\sim$ hundreds of MHz), LOFAR and GMRT surveys confirm that FR0s typically lack the large-scale, optically thin synchrotron emission characteristic of classical FRI/II RGs. The majority exhibit flat or slightly convex radio spectra, which contrasts with the highly peaked spectra typical of young RGs and reinforces the idea that FR0s do not possess significant extended or infant jet structures \citep{capetti19,capetti20b,ishwara20,chilufya24}.

At GHz frequencies, most FR0s have flat spectra and and appear unresolved or barely unresolved on kpc scales \citep{chhetri13,sadler14,whittam20,mikhailov21b}. Dedicated VLA observations achieving sub-mJy sensitivities show that the majority of FR0s still remain unresolved ($\mathbf{<}$ a few kpc, Fig.~\ref{diagram}). Crucially, only $\sim$30\% of FR0s display low-brightness, symmetric kpc-scale jets, suggesting the possible presence of substantial extended emission in some cases that is detectable at low frequencies \citep{baldi15,baldi19}.

Intermediate-resolution eMERLIN observations at 5 GHz ($\sim$ tens of mas) have detected weak pc-scale core emission, though the core contributes to only a small fraction ($<$20\%) of the total flux \citep{baldi21c}. This result highlights the critical importance of including long baselines ($>$ hundreds of km) in isolating compact, self-absorbed, flat-spectrum cores from surrounding steep-spectrum extended jets at high angular resolution.

The highest-resolution (fraction of mas) studies using VLBI technique (VLBA, EVN) have revealed resolved, symmetric pc-scale jets in about 80\% of FR0s \citep{cheng18,cheng21,giovannini23}. These jets display low jet-to-counter-jet ratios, indicating mildly relativistic bulk speeds. These observations confirm that the lack of extended emission is not solely due to observational limits but also reflects intrinsic jet properties, such as reduced jet stability or production efficiency. Combining VLA-eMERLIN datasets has proven effective for resolving intermediate-scale jets ($<$ 1 kpc, \citealt{baldi21c}). This multi-scale approach emphasizes the need for more extensive VLBI observations to fully characterize the jet physics of FR0s \citep{baldi24} and to extend their exploration to high redshifts.

\section{The role of SKA in investigating FR0s}

The physical nature of FR0 RGs remains elusive due to current limitations in sensitivity and spatial resolution. The forthcoming SKA design baseline, Array Assembly 4 (AA4), will be decisive in addressing these uncertainties by providing high-fidelity imaging, exceptional sensitivity, and statistically complete sample studies through both large-scale surveys and VLBI follow-ups.

The SKA-Mid AA4 configuration will offer sub-$\mu$Jy sensitivities and sub-arcsecond resolution. With continuum sensitivities down to $\sim$0.25 $\mu$Jy beam$^{-1}$ in 8 hours at 6 GHz, SKA-Mid is uniquely suited to detect and characterize the compact, weak radio cores of FR0s at z $<$ 0.1 down to luminosities $< 10^{19}$ W Hz$^{-1}$ with $0.14\arcsec$ resolution ($\sim 250$ pc, Briggs robust $= 0$). SKA-Low will complement this by tracing low-frequency synchrotron emission with $\sim 12$ $\mu$Jy beam$^{-1}$ sensitivity in 1 hour (reaching the confusion limit), enabling the detection of faint diffuse structures with $10\arcsec$ angular resolution at 200 MHz. In VLBI mode, SKA-Mid can be phased within global networks (e.g., EVN, VLBA, LBA, leveraging North-South and East-West baselines), providing $\sim 1$ mas ($\sim 2$ pc at $z \sim 0.1$) imaging at 5 GHz with an rms noise of a few tens of $\mu$Jy beam$^{-1}$ in hours. This is crucial for distinguishing high brightness temperature ($T_{B} > 10^{6} \text{ K}$) AGN cores from star formation (SF) \citep{morabito22}. SKA-Low VLBI will extend these capabilities to the MHz regime, and commensal SKA-VLBI observations will maximize scientific output by leveraging, for instance, European stations to fully exploit their observing time, greatly expanding both the discovery and the characterization of new FR0 samples.

\textit{FR0 selection}. The definition of the FR0 class is not conclusive because both resolution and sensitivity are critical for distinguishing compact sources from extended RGs and from other compact analogues such as blazars, RQAGN, starburst galaxies and young RGs. A robust selection requires careful consideration of both radio and optical properties, alongside the technical limitations of the utilized surveys. The first systematic attempt to define and select FR0s was made by \citet{baldi15, baldi18}, who introduced the term for local ($z < 0.05$) compact RLAGN with the following characteristics: i) a compact radio morphology, where sources appear unresolved or marginally resolved in arcsecond-scale radio surveys (on galaxy scale, $\lesssim 10$ kpc); ii) a low radio flux density cut (e.g., $> 5$ mJy at 1.4 GHz in FIRST, \citealt{becker95}) to ensure sufficient signal-to-noise ratio ($>$ 10) for classification and to avoid contamination from star-forming galaxies, which dominate at the sub-mJy level \citep{bonzini13}; iii) a nuclear optical spectroscopic classification as LERGs is essential, using emission-line diagnostics \citep{kewley06} to confirm the radio emission is not SF-driven; iv) a selection based on host galaxy properties to favor red massive early-type galaxies ($M_* \gtrsim 10^{10} M_\odot$), which are more likely to host RLAGN \citep{best05b, sabater19}. Future FR0 selection will benefit from next-generation deep, high-resolution radio and optical/near-infrared surveys. 

\textit{Population studies.} SKA-Mid/Low campaigns will scan a vast number of nearby galaxies with $\mu$Jy beam$^{-1}$ sensitivity, enabling direct morphological classification across a wider luminosity and redshift range and improving the detection of faint extended emission. A SKA survey reaching $\sim$1 $\mu$Jy beam$^{-1}$ at $\sim$1 GHz over 30 deg$^{2}$ will detect $\sim$3 $\times 10^{5}$ low-power AGN \citep{prandoni15,padovani16}. Dedicated and commensal SKA-VLBI observations will provide deeper follow-up studies of FR0 samples (down to sub-mas scale at some GHz), detected by deep SKA-Mid/Low campaigns. These statistics, crucial for understanding the dominant faint radio emitters, will enable the community to perform population-level studies on BH activity and BH-host co-evolution for low-luminosity radio AGN in general throughout cosmic history.

\section{SKA deliverables}



SKA will be revolutionary for studying faint, compact radio structures and will shed new light on the following key aspects of the FR0 RGs in particular: accretion-ejection mechanisms, environment and feedback, identification of high-redshift populations, and the duty cycle of RGs.

\subsection{Accretion-ejection}

The compactness of FR0s is explained by two competing scenarios. The {\bf engine} or {\bf static scenario} attributes it to intrinsic limitations in jet launching, potentially due to low BH spin, weak magnetic fields, or jet instability \citep{garofalo19,baldi23}. In this view, FR0s' lack of extended structures is a fundamental property of the central engine. Conversely, the {\bf dynamic scenario} posits that FR0s are a transient phase in evolution, either as young sources that have not yet developed extended jets or as systems with recurrent, short-lived episodes of jet activity \citep{sadler14, capetti19, odea21}.

To distinguish between these scenarios, SKA-VLBI will provide constraints on several fundamental physical aspects of the FR0s: i) the efficiency of jet production (extended versus unresolved emission), linked to their low accretion rates \citep{amarantidis25}, and energy dissipation during propagation \citep{rossi20} through morphological and polarization studies; ii) the role of BH spin and magnetic flux by resolving jet structures near the event horizon for nearby FR0s with global VLBI ($\sim$0.5 mas at 15 GHz) and using the core-shift technique \citep{chatterjee25}; iii) jet collimation profiles and opening angles, which also depend on BH spin \citep{tchekhovskoy11,krause12}; iv) within the jet-launching region, the transverse width profile to distinguish between parabolic and conical shapes, which correspond to different jet models (e.g.. \citealt{blandford77,nemmen14,boccardi21,baan25}).

Regarding jet propagation, SKA observations with long and short baselines and large $uv$ coverage can resolve the structure of jets, testing the emissivity of a weak central spine with respect to a slower dominant outer layer in comparison with the FRI stratified jet model \citep{ghisellini05} and distinguishing FR0s from other compact radio impostors (e.g., by considering beaming effects, diffuse emission, and brightness temperature). The high image fidelity is crucial for correct jet positioning to enable  kinematic studies to constrain the bulk speed and for polarization studies to probe the magnetic field strength at parsec scale, providing insight into why the extended flows fail. The combination of observations with short and long baselines  will also allow to fill up the u-v plane to consistently image the core and outer jet parts. Combining continuum, rotation measure, and polarized maps will help to determine whether FR0 jets develop instabilities \citep{costa24,costa26,lalakos24,boula25,cardoso26}, which may prevent or allow the formation of an ordered poloidal magnetic structure at large distance from the core.

Multi-frequency SKA observations at sub-arcsec resolutions can instead study the large-scale jet structure. Jet sidedness, jet prominence, and core dominance can constrain the jet bulk Lorentz factors and viewing angle, assuming a randomly oriented population of RGs. Linear sizes and spectral index maps help trace energy losses, ages, particle content, and acceleration/deceleration within a spine-layer jet model \citep{webster21b,boughelilba23}.

Finally, SKA will constrain the timescales of radio outbursts and the typical lifetimes of FR0s \citep{jurlin20}. With broad spectral coverage, SKA-Mid and SKA-Low will test whether (and estimate their fraction) they are genuinely short-lived young sources ($<$10$^{5}$ yr) with peaked radio spectra \citep{odea21} and display remnants of past activity ($>$ 10$^{6}$ yr) on a large scale \citep{capetti19,shabala20}. Since their number density strongly suggests they do not generally evolve into large RGs \citep{baldi19}, SKA-Low observations, in particular, will search for ultra-steep spectrum relic emission, which would constrain the most recent past ejection in the case of intermittent activity \citep{capetti20}. SKA-Mid/Low surveys and VLBI follow-ups will allow statistical population studies regarding the morphological compactness and jet alignment with respect to the galaxy disk to test the role of the BH spin axis tilt in causing a frustrated or precessing jet \citep{liska18,stanghellini25}.

\subsection{Environment and feedback}

In the past, the role of low-luminosity RLAGN, including FR0s, in galaxy evolution has often been underestimated compared to powerful radio sources, which by definition can offer a larger energetic budget to the surroundings. Currently, deep radio surveys are reversing this view, showing that these low-power jets continuously inject energy into the host galaxy's ISM \citep{sabater19}. In contrast to powerful jets that "drill" through the ISM, low-power jets are  also susceptible to galaxy-scale disruption and entrainment, which increases the volume and duration of feedback and energy transfer to the ISM \citep{mukherjee16,mukherjee18a}. These findings have major implications for understanding radio-mode feedback and the co-evolution of BHs and galaxies at low luminosity regimes \citep{fabian12}. Recent observations of X-ray cavities and gas disturbances reveal that compact jets  can drive effective feedback by heating the ISM and suppressing SF, supporting the idea that even low-power ejecta are crucial for galaxy-BH self-regulation \citep{allen06,balmaverde08,nesvadba07,ubertosi21}.

The surrounding gas (the environment in general) can also characterize FR0 evolution and radio-mode feedback. While FR0s and FRIs have similar host poor-gas properties \citep{baldi18} with possible jet entrainment for stellar mass-loading (e.g. \citealt{perucho14,oshea25}), optical studies show FR0s typically reside in moderately rich Mpc-scale clusters \citep{vardoulaki21}, but slightly less dense than those of FRIs \citep{capetti20b}. This suggests the environment may play a role in limiting jet development, although it is likely not the sole factor. Synergies with other optical/IR/X-ray facilities will be key to addressing this question.

Several key questions about FR0 feedback and environment remain open. Firstly, how do FR0 jets interact with their ISM? SKA VLBI's high-resolution imaging (continuum and polarization) can be combined with multi-wavelength (photometry and spectroscopy) data from optical ground-based/space telescopes and millimeter/sub-millimeter telescopes to provide an unprecedented view of jet-ISM interaction (position and physics). This can reveal signatures of mechanical feedback, such as localized changes in brightness or gas kinematics in different phases, allowing for a quantitative estimate of the energy released \citep{ruffa20,venturi21}. Secondly, what is the energetic budget released by FR0s? SKA-Low's deep, low-frequency observations, possibly combined with X-ray observations, are crucial here because they can trace the long-lived synchrotron-aged plasma from past jet activity, which inflated  X-ray cavities \citep{cavagnolo10}. This provides insight into the integrated feedback history and duty cycle of FR0s, revealing the true extent of jet expansion into the galaxy and allowing a better characterization of their jet physical properties. Finally, what is the role of the environment for FR0s? SKA radio continuum and polarization observations, combined with multi-wavelength data on gas content (e.g., hot gas, HI, H$_{2}$, CO lines), will allow us to assess how the environment (e.g., dark matter halo, clustering, ISM) and BH accretion shape RLAGN populations and affect jet confinement and efficiency (e.g., \citealt{ruffa19a,croston19,grandi21}). This will be critical for determining whether FR0s are intrinsically jet-inefficient and how strong feedback results from their jet suppression.

\subsection{High-redshift FR0s and duty cycles of LERG population}

The LERG population (including FR0s, FRIs, and possibly a sub-class of FRIIs) is believed to belong to the same parent population of massive early-type galaxies hosting low-accreting BHs \citep{baldi23,ye25,grandi25}. In this context, FR0s may represent a specific stage of the cosmic evolution of LERGs, triggered by a combination of factors like BH spin, magnetic fields, and environment \citep{grandi21,kondapally22}. The emergence of FR0s as the most numerous RLAGN class has fundamentally reshaped our understanding of the LERG population in the local Universe. However, their high-redshift counterparts remain poorly constrained. There is growing evidence for compact RGs at $z > 1$ in deep surveys \citep{baldi13b,smolcic17,bondi18, williams18, vardoulaki21}, but current facilities lack the sensitivity and resolution to reliably distinguish them from other sources, such as star-forming galaxies.

SKA-Mid will transform this picture, with sensitivities down to $\mu$Jy, sub-arcsecond resolution and broad spectral coverage, enabling the detection of compact radio sources out to z  $\gtrsim$ 2 at sub-galactic scales ($\gtrsim$ 1 kpc) down to L $\sim 10^{22} \text{ W Hz}^{-1}$. Concurrently, SKA-Low will be sensitive to the diffuse, steep-spectrum emission from unresolved cores or aged lobes, revealing signs of past AGN activity. By constructing redshift-dependent luminosity functions, SKA campaign will robustly separate radiatively inefficient AGN (LERGs, mostly FR0/I) from other AGN/SF-dominated populations and track the BH activity of the most massive galaxies in the Universe \citep{prandoni18,delvecchio22}. Then, dedicated VLBI studies will identify and confirm compact cores and resolve sub-kpc jets, allowing the exploration of FR evolution across cosmic time and the statistical determination of AGN duty cycles \citep{kondapally23}. For example, comparing local and distant samples can test if FR0s represent brief, re-triggered phases of AGN activity. This will ultimately determine if the compact, core-dominated AGN mode is a recent, low-$z$ phenomenon or a persistent state throughout cosmic history \citep{jackson99}.

Two primary open questions remain. First, are FR0s the most abundant RLAGN population at cosmic noon? SKA will cross-match radio-selected FR0s with optical-infrared surveys to determine host galaxy demographics at z $\sim$ 1-2. This will clarify the balance between RL and RQ AGN populations across cosmic time, testing if FR0s account for most RL activity during the peak epoch of galaxy growth. Second, is the role of FR0s at high $z$ crucial for feedback? At cosmic noon, galaxies are more actively star-forming and reside in denser environments \citep{carilli13}. In these conditions, the compact jets of FR0s may interact more effectively with the ISM, potentially making them more efficient regulators of galaxy growth than their extended FRI/II relatives \citep{nesvadba07, mukherjee16,borodina25}.

\section{Conclusions and multi-band synergy perspectives}

Since the FR0 phase represents a common, low-power mode of jet-driven feedback that likely plays a dominant role in regulating SF and gas cooling in both the local and distant Universe, the $\mu$Jy sensitivity, sub-arcsecond resolution, and broad-band coverage of SKA, including both survey and VLBI capabilities, will be crucial for investigating the co-evolution of BHs and galaxies at low luminosities. However, the study of FR0s fundamentally requires a multi-wavelength and multi-scale approach. While radio observations are indispensable, a comprehensive understanding of their accretion-ejection processes, host galaxies, environments, and feedback mechanisms depends on integrating data across the electromagnetic spectrum. The unique SKA extragalactic continuum survey and VLBI capabilities will synergize powerfully with existing and forthcoming facilities in the sub/millimeter (e.g., ALMA, AtLAST, IRAM, ngVLA), optical/infrared (e.g., JWST, HST, VLT, Euclid, LSST), X-ray (Chandra, XMM, NewAthena, AXIS), and $\gamma$-ray (Fermi-LAT, LHAASO, CTAO) bands, as well as with neutrino detectors (IceCube, KM3NeT) via both large surveys and targeted follow-up instruments. This combined multi-band effort will enable a transformative, extinction-free census of active BHs across a wide range of galaxy types, environments, and redshifts.

\bibliographystyle{abbrvnat-maxbibnames4}
\bibliography{my} 

\end{document}